\documentclass[12pt]{article}
\usepackage[english]{babel}
\usepackage{graphicx,epsfig}
\makeindex \textwidth 170mm \textheight 220mm \topmargin -5mm
\newcommand{\be}{\begin{equation}}
\newcommand{\ee}{\end{equation}}
\newcommand{\bea}{\begin{eqnarray}}
\newcommand{\eea}{\end{eqnarray}}

\def\derp#1#2{\rp{\partial{#1}}{\partial{#2}}}
\def\rfr#1{eq.(\ref{#1})}
\def\rfrs#1#2{eqs.(\ref{#1})-(\ref{#2})}

\def\Rfrs#1#2{Eqs.(\ref{#1})-(\ref{#2})}
\def\derp#1#2{\rp{\partial{#1}}{\partial{#2}}}
\def\dert#1#2{\frac{{{d}}{#1}}{{{d}}{#2}}}
\def\bb{\bibitem}
\def\eqi{\begin{equation}}
\def\eqf{\end{equation}}
\def\eqia{\begin{eqnarray}}
\def\eqfa{\end{eqnarray}}

\def\rp#1#2{{#1\over#2}}

\def\lb#1{\label{#1}}

\def\bm#1{{\mbox{\boldmath$#1$\unboldmath}}}
\def\parp{\left(1+\trt\right)}
\def\parm{\left(1-\trt\right)}

\def\trt{\frac{k\cos\theta}{r^2}}

\begin{document}
\begin{titlepage}
\begin{flushright}
\today\\
BARI-TH/00\\
\end{flushright}
\vspace{.5cm}
\begin{center}
{\LARGE Some comments on a recently derived approximated solution
of the Einstein equations for a spinning body with negligible
mass} \vspace{1.0cm}
\quad\\
{Lorenzo Iorio$^{\dag}$\\
\vspace{0.1cm}
\quad\\
{\dag}Dipartimento di Fisica dell' Universit{\`{a}} di Bari, via
Amendola 173, 70126, Bari, Italy\\ \vspace{0.1cm} }
\quad\\
 \vspace{1.0cm}

{\bf Abstract\\}
\end{center}

{\noindent \footnotesize Recently, an approximated solution of the
Einstein equations for a rotating body whose mass effects are
negligible with respect to the rotational ones has been derived by
Tartaglia. At first sight, it seems to be interesting because both
external and internal metric tensors have been consistently found,
together an appropriate source tensor; moreover, it may suggest
possible experimental checks since the conditions of validity of
the considered metric are well satisfied at Earth laboratory
scales. However, it should be pointed out that reasonable doubts
exist if it is physically meaningful because it is not clear if
the source tensor related to the adopted metric can be realized by
any real extended body. Here we derive the geodesic equations of
the metric and analyze the allowed motions in order to disclose
possible unphysical features which may help in shedding further
light on the real nature of such approximated solution of the
Einstein equations.}
\end{titlepage} \newpage \pagestyle{myheadings} \setcounter{page}{1}
\vspace{0.2cm} \baselineskip 14pt

\setcounter{footnote}{0}
\setlength{\baselineskip}{1.5\baselineskip}
\renewcommand{\theequation}{\mbox{$\arabic{equation}$}}
\noindent \section{Introduction} As it is well known, the Kerr
metric is a stationary solution of the Einstein field equations in
the vacuum which is endowed with axial symmetry and is
characterized by two independent parameters: the asymptotic mass
and the asymptotic angular momentum per unit mass. Unfortunately,
it would probably be incorrect to adopt such metric in order to
describe the motion of light rays and test masses in the
gravitational field of any real material rotating body because, at
present, the problem of finding an internal mass-energy
distribution corresponding to the Kerr metric has not yet received
a complete general answer. Many efforts have been dedicated to the
construction of sources which could represent some plausible
models of stars. In \cite{hartho68} the authors have tried to
connect the external Kerr metric to the multipolar structure of
various types of stars assumed to be rigidly and slowly rotating.
Terms of greater than the second order in the angular velocity
were neglected. In \cite{ali01} a physically reasonable fluid
source for the Kerr metric has been obtained. For previous
attempts to construct sources for the Kerr metric see
\cite{hagmar81, hag90}. In \cite{wil03} a general class of
solutions of Einstein's equations for a slowly rotating fluid
source, with supporting internal pressure, is matched to the Kerr
metric up to and including first order terms in angular speed
parameter. In \cite{bicled93} counterrotating thin disks of finite
mass, consisting of two streams of collisionless particles,
circulating in opposite directions with different velocities are
examined. The authors show how such disks can act as exact sources
of all types of the Kerr metric.

Recently, in \cite{tart03} an approximated solution of the
Einstein field equations for a rotating, weakly gravitating body
has been found. It seems promising because both external and
internal metric tensors have been consistently found, together an
appropriate source tensor. Moreover, the mass and the angular
momentum per unit mass are assumed to be such that the mass
effects are negligible with respect to the rotation effects: this
is a situation which is rather easy to obtain, especially at
laboratory scale \cite{tart02}. As a consequence, only quadratic
terms in the body's angular velocity has been retained. One main
concern is that it is not clear if there is a real mass
distribution able to generate the found source tensor.

In this paper we wish to derive some consequences which could
allow to shed light on the physical nature of the considered
metric. Of course, if some exotic effects in apparent contrast
with the experience will be found, this would represent a further,
strong sign that the examined metric is not physically reasonable
in describing a real rotating source. There are well developed
theoretical methods to exclude pathological metrics from further
considerations, but we think that fully deriving some consequences
which may turn out to be unphysical  is less formal and more
vivid.

The paper is organized as follows. In Section 2 we briefly review
the metric. In Section 3 we calculate the geodesic equations of
the metric both in spherical and in cartesian coordinates and
consider free fall and weighing scenarios in a terrestrial
laboratory context. Section 4 is devoted to the conclusions.
\section{An approximated solution of the Einstein equations for a weakly gravitating, spinning body}
Here we wish to investigate the case of a rotating body of mass
$M$, radius $R$ and proper angular momentum $J$ with negligible
mass with respect to the rotation in the sense that \eqi
a=\frac{J}{Mc}>>R_{\rm s}=\frac{2GM}{c^2},\lb{condiz}\eqf where
$G$ is the Newtonian gravitational constant. This is an
interesting situation which is quite common at laboratory scales
and also in some astronomical situations \cite{tart02}. For, say,
a rotating sphere with\footnote{These critical values for the
mechanical parameters of the sphere can be obtained with the
cutting-edge technologies available today or in the near future
\cite{foretal94}, cited in \cite{tart02}.} $\omega=4.3\times 10^4$
rad s$^{-1}$, $R=2.5\times 10^{-2}$ m, $M=1.11\times 10^{-1}$ kg
we would have $a=3.6\times 10^{-8}$ m and $R_{\rm s}=1.6\times
10^{-28}$ m, while the Earth has $a_{\oplus}=3.3$ m and $R_{\rm
s}=8.86\times 10^{-3}$ m. An approximated solution of the Einstein
equations for a body which satisfies the condition of \rfr{condiz}
has been recently obtained in \cite{tart03}.

In a frame whose origin coincides with the center of the spinning
mass let us adopt the spherical coordinates $(r,\ \theta,\ \phi)$
with the $\theta$ angle counted from the axis of rotation of the
body; then, the external solution is \cite{tart03} \eqi (ds)^2\sim
c^2(1+h_{00})(dt)^2-(1+h_{rr})(dr)^2-r^2(1+h_{\theta\theta})(d\theta)^2-
r^2\sin^2\theta(1+h_{\phi\phi})(d\phi)^2,\lb{linel}\eqf with
\bea h_{00}&=& C_0\frac{a^2}{r^2}\cos\theta,\lb{uno}\\
h_{rr}=h_{\theta\theta}=h_{\phi\phi}&=&-C_0\frac{a^2}{r^2}\cos\theta,\lb{due}\eea
where, for a rotating homogeneous sphere,
$a=\frac{2}{5}\frac{\omega R^2}{c}$ and the value
$C_0=\frac{50}{3}\pi$ can be obtained only on the base of an
analogy \cite{tart03}; however, it should be of the order of
unity. \Rfrs{uno}{due} has been obtained by assuming
$\varepsilon=\frac{R_{\rm s}}{2r}\ll \alpha=\frac{a}{r}.$ This is
the reason why the off--diagonal gravitomagnetic term, linear in
$\omega$, has been neglected.

The interesting point is that, contrary to the Kerr solution which
cannot be extended to the interior of a generic material body, it
has an internal counterpart. On the other hand, it is questionable
if such metric is really able to describe any material body:
indeed, the source tensor for a rotating homogeneous sphere
yielded by the Tartaglia's metric is \eqia T_0^0 &=&
-\frac{6\omega^2}{25\pi
c^2}C_0\cos\theta,\lb{tmunuoo}\\
T_r^r=T_{\theta}^{\theta}=T_{\phi}^{\phi}&=&0.\lb{tmunuff}\eqfa
Note that $T_0^0$ is antisymmetric for reflections with respect to
the equatorial plane and its integral over the entire volume of
the sphere vanishes. Now the question is: what kind of real mass
distribution could give rise to such an energy-momentum tensor as
that of \rfrs{tmunuoo}{tmunuff}? Note also that an elastic
energy-momentum tensor should be added to it in order to account
for the elastic force needed to keep the whole body together
against the centrifugal forces and, from a mathematical point of
view, to insure the continuity of the radial derivatives of the
metric tensor at the boundary of the body. The metric that
Tartaglia gives does not contain the influence of such additional
parts of the energy-momentum tensor and its significance must
therefore be investigated in terms of the source that it does
represent.

The observable consequences derived in the following section could
shed some light to this problem.

\section{The geodesic motion of a test particle}
Let us investigate the free motion of a point mass in the metric
of \rfrs{uno}{due}.

By defining $k=C_0 a^2=\frac{4}{25}C_0\frac{\omega^2 R^4}{c^2}$,
from \rfrs{uno}{due} the determinant $g$ of the metric tensor
$g_{\mu\nu}$ is \eqi
g=-r^4\sin^2\theta\left(1+\trt\right)\left(1-\trt\right)^3.\eqf
Then, the inverse of $g_{\mu\nu}$ is \eqia g^{00} &=&
\frac{1}{\parp},\lb{g00}\\
g^{rr}&=& -\frac{1}{\parm},\\
g^{\theta\theta}&=& -\frac{1}{r^2 \parm},\\
g^{\phi\phi} & = & -\frac{1}{r^2\sin^2\theta\parm}. \lb{gff}\eqfa
The geodesic equations of the motion of a test particle are \eqi
\frac{d^2
x^{\mu}}{d\tau^2}+\Gamma^{\mu}_{\nu\rho}\dert{x^{\nu}}{\tau}\dert{x^{\rho}}{\tau}=0,\eqf
where $\tau$ is the proper time of the particle and the
$\Gamma^{\mu}_{\nu\rho}$ are the Christoffel symbols
\eqi\Gamma^{\mu}_{\nu\rho}=\frac{g^{\mu\alpha}}{2}\left(\derp{g_{\alpha\nu}}{x^{\rho}}+
\derp{g_{\alpha\rho}}{x^{\nu}}-\derp{g_{\nu\rho}}{x^{\alpha}}\right).\eqf
From \rfrs{uno}{due} and \rfrs{g00}{gff} it turns out that the
only non vanishing Christoffel symbols are \eqia \Gamma^0_{0r} & =
&-\frac{k\cos\theta}{r^3\parp},\\
\Gamma^0_{0\theta} &=& -\frac{k\sin\theta}{2r^2\parp},\\
 \Gamma^{r}_{00}
&=& -\frac{k\cos\theta}{r^3\parm},\\
 \Gamma^{r}_{rr}
&=& \frac{k\cos\theta}{r^3\parm},\\
\Gamma^{r}_{\theta\theta}
&=& -\frac{r}{\parm},\\
\Gamma^{r}_{\phi\phi}
&=& -\frac{r\sin^2\theta}{\parm},\\
\Gamma^{r}_{r\theta}
&=& -\frac{k\sin\theta}{2r^2\parm},\\
\Gamma^{\theta}_{00}
&=& -\frac{k\sin\theta}{2r^4\parm},\\
\Gamma^{\theta}_{rr}
&=& -\frac{k\sin\theta}{2r^4 \parm},\\
\Gamma^{\theta}_{\theta\theta}
&=& \frac{k\sin\theta}{2r^2\parm},\\
\Gamma^{\theta}_{\phi\phi}
&=& \frac{-2r^2\sin\theta\cos\theta+k\sin\theta(3\cos^2\theta-1)}{2 r^2 \parm},\\
\Gamma^{\theta}_{r\theta}
&=& \frac{1}{r\parm},\\
\Gamma^{\phi}_{r\phi} &=& \frac{1}{r\parm},\\
\Gamma^{\phi}_{\theta\phi} &=&
\frac{2r^2\sin\theta\cos\theta-k\sin\theta(3\cos^2\theta-1)}{2
r^2\sin^2\theta\parm}. \eqfa Then, the geodesic equations for $t,\
r,\ \theta$ and $\phi$ are, in explicit form\footnote{Of course,
they can also be derived from the Lagrangian
$\mathcal{L}=\frac{m}{2}g_{\mu\nu}{\dot x^{\mu}}{\dot x^{\nu}}$
and the Lagrange equations $\dert{\left(\derp{{\mathcal{L}}}{\dot
x^{\mu}}\right)}{\tau}-\derp{{\mathcal{L}}}{x^{\mu}}=0$. } \eqia
\ddot{t}\parp &=& \frac{2k\cos\theta}{r^3}\dot t\dot r+\frac{k\sin\theta}{r^2}\dot t\dot\theta,\lb{eqdifft}\\
\ddot{r}\parm
&=&r(\dot\theta)^2+r\sin^2\theta(\dot\phi)^2+\frac{c^2
k\cos\theta}{r^3}(\dot t)^2-\frac{k\cos\theta}{r^3}(\dot
r)^2-\frac{k\sin\theta}{r^2}\dot
r\dot\theta,\lb{geodr}\\
\ddot{\theta}\parm &=& -\frac{2\dot\theta\dot r
}{r}+\sin\theta\cos\theta(\dot\phi)^2+\frac{c^2k\sin\theta}{2r^4}(\dot
t)^2 +\frac{k\sin\theta}{2r^4}(\dot r)^2-\nonumber\\\lb{geodtheta}
&-&\frac{k\sin\theta}{2r^2}(\dot \theta)^2
-\frac{k\sin\theta(3\cos^2\theta-1)}{2r^2}(\dot\phi)^2,\\
\ddot{\phi}\parm &=&-\frac{2\dot
r\dot\phi}{r}-\frac{2\cos\theta}{\sin\theta}\dot\theta\dot\phi+\left[\frac{k}{r^2\sin\theta}(3\cos^2\theta-1)\right]
\dot\theta\dot\phi.\lb{geodphi}\eqfa Note that in such equations
the terms $\left(1\pm\frac{k\cos\theta}{r^2}\right)$ on the
left-hand sides can be considered almost equal to 1 because
$\frac{k}{r^2}$ is of the order of $10^{-11}$ or less. For the
same reason, since
$d\tau=\sqrt{g_{00}}dt=\sqrt{1+\frac{k\cos\theta}{r^2}}dt$ we can
safely assume $\dot t=1$. Note also that for $k=0$
\rfrs{geodr}{geodphi} reduce to the equations of motion in a flat
spacetime in spherical coordinates, i.e.
\eqia a_{r}\equiv\ddot{r}-r(\dot\theta)^2-r\sin^2\theta(\dot\phi)^2&=&0,\\
a_{\theta}\equiv r\ddot{\theta}+2\dot\theta\dot r
-r\sin\theta\cos\theta(\dot\phi)^2&=&0,\\
a_{\phi}\equiv r\sin\theta\ddot{\phi}+2\sin\theta\dot
r\dot\phi+2r\cos\theta\dot\theta\dot\phi&=&0, \eqfa where $a_{r},\
a_{\theta},\ a_{\phi}$ are the three components of the particle's
acceleration in spherical coordinates.


Let us analyze some possible types of motion for a point particle
obeying \rfrs{geodr}{geodphi}, which are, as it can be seen,
highly nonlinear and difficult to resolve without some simplifying
assumptions. A purely radial motion is not allowed. Indeed, for
$\theta$ and $\phi$ constant \rfr{geodtheta} yields \eqi c^2+(\dot
r)^2=0\eqf which does not admit solution.

Consider now the case of a motion with $r=r_0$ and
$\theta=\theta_0$. The geodesic equations reduce to

\eqia \ddot t\left(1+\frac{k\cos\theta_0}{r_0^2}\right)&=&0,\lb{primaa}\\
\frac{kc^2\cos\theta_0}{r_0^3}(\dot t)^2+r_0\sin^2\theta_0 (\dot
\phi)^2&=&0,\lb{prima}\\
\frac{kc^2}{r_0^2}(\dot t)^2+[-2r_0^2\cos\theta_0+k(3\cos^2\theta_0-1)](\dot\phi)^2&=&0,\lb{seconda}\\
\ddot\phi\left(1-\frac{k\cos\theta_0}{r_0^2}\right)&=&0.
\lb{terza}\eqfa For $\theta_0=\frac{\pi}{2}$ \rfr{prima} yields
$r_0(\dot\phi)^2=0$, i.e. $\phi=\phi_0=$ constant. This condition
satisfies \rfr{terza} but, from \rfr{seconda}, it would imply
$\frac{kc^2}{r^2_0}=0$. So, the  equations for $r=r_0,\
\theta=\theta_0=\frac{\pi}{2}$ do not admit solutions.

Would it be possible a circular motion along a parallel for
$r=r_0$ and $\theta=\theta_0\neq\frac{\pi}{2}$? If
$\cos\theta_0>0$, \rfr{prima} can never be  satisfied. On the
other hand, for $\cos\theta_0<0$ \rfr{prima} can be satisfied,
while \rfr{seconda} does not admit any solution because it can be
shown that $2r_0^2\cos\theta_0+k(1-3\cos^2\theta_0)> 0$ cannot be
satisfied for $\frac{\pi}{2}<\theta_0<\pi$. This means that
motions with $r$ and $\theta$ constant cannot occur even outside
the equatorial plane of the source.

The motion along a meridian with $r=r_0$ and $\phi=\phi_0$ is
governed by the equations \eqia
\ddot{t}\left(1+\frac{k\cos\theta}{r_0^2}\right) -\frac{k\sin\theta}{r_0^2}\dot t\dot\theta &=& 0,\lb{one}\\
\frac{c^2 k\cos\theta}{r_0^3}(\dot
t)^2+r_0(\dot\theta)^2 &=& 0,\lb{two}\\
\ddot{\theta}\left(1-\frac{k\cos\theta}{r_0^2}\right)
-\frac{c^2k\sin\theta}{2r_0^4}(\dot t)^2
+\frac{k\sin\theta}{2r_0^2}(\dot \theta)^2 &=& 0.\lb{three}\eqfa
The geodesic equation for $\phi$ identically vanishes. From
\rfr{two} it follows that a motion for $0<\theta<\frac{\pi}{2}$,
i.e. $\cos\theta>0$, is not allowed.

The "spherical" motion with $r=r_0$ is described by
\eqia\ddot t\left(1+\frac{k\cos\theta}{r_0^2}\right)&=& \frac{k\sin\theta}{r_0^2}\dot t\dot\theta,\lb{one1}\\
\frac{c^2 k\cos\theta}{r_0^3}(\dot
t)^2&=&-r_0(\dot\theta)^2 -r_0\sin^2\theta(\dot\phi)^2,\lb{two2}\\
\ddot{\theta}\left(1-\frac{k\cos\theta}{r_0^2}\right)&=&
+\frac{c^2k\sin\theta}{2r_0^4}(\dot t)^2
-\frac{k\sin\theta}{2r_0^2}(\dot \theta)^2 -\nonumber\\
&-&\frac{\sin\theta}{2r_0^2}\left[-2r_0^2\cos\theta+k(3\cos^2\theta-1)\right](\dot\phi)^2\lb{three3},\\
\ddot\phi\left(1-\frac{k\cos\theta}{r_0^2}\right)&=&\left[-\frac{2\cos\theta}{\sin\theta}-
\frac{k}{r_0^2\sin\theta}(3\cos^2\theta-1)\right]
\dot\theta\dot\phi.\eqfa From \rfr{two2} it follows that a motion
for $0<\theta<\frac{\pi}{2}$, i.e. $\cos\theta>0$, is not allowed.
Finally, after such rather pathological situations, let us
investigate the case $r=r_0,\ \theta=\theta_0, \phi=\phi_0, \ \dot
r=0,\ \dot\theta=0, \dot\phi=0$. We have \eqia
\ddot t &=& 0,\\
\ddot r &=& \frac{kc^2\cos\theta_0}{r_0^3},\\
\ddot\theta &=& \frac{kc^2\sin\theta_0}{2r_0^4},\\
\ddot\phi &=&0.\eqfa Such results can describe the relativistic
acceleration experienced by a material sample suspended over the
rotating central body. For example, for $\theta_0=0$ we have that
the sample would be acted upon by an acceleration of magnitude
$\frac{kc^2}{r_0^3}$ directed upwards along the local vertical
which would tend to counteract the Earth's gravitational
acceleration $g$ in a laboratory experiment. This topic will be
treated in more details in the following subsection.

In order to derive some more vivid consequences of the geodesic
equations of motion and visualize them it is helpful to adopt the
cartesian coordinates.

By using \eqia dr &=&
dx\sin\theta\cos\phi+dy\sin\theta\sin\phi+dz\cos\theta,\\
rd\theta &=&
dx\cos\theta\cos\phi+dy\cos\theta\sin\phi-dz\sin\theta,\\
r\sin\theta d\phi &=& -dx\sin\phi+dy\cos\phi,
\\\cos\theta&=&\frac{z}{r}\eqfa in the metric in spherical coordinates it is
possible to express the $g_{\mu\nu}$ in cartesian coordinates
\eqia
h_{00}&=&\frac{kz}{(x^2+y^2+z^2)^{\frac{3}{2}}},\lb{cartesian1}\\
h_{11}=h_{22}=h_{33}&=&-\frac{kz}{(x^2+y^2+z^2)^{\frac{3}{2}}}.\lb{cartesian2}\eqfa

From \rfrs{cartesian1}{cartesian2} the Lagrangian of a particle
with mass $m$ is
\eqi\mathcal{L}=\frac{m}{2}\left\{\left[1+\frac{kz}{(x^2+y^2+z^2)^{\frac{3}{2}}}\right]c^2(\dot
t)^2
-\left[1-\frac{kz}{(x^2+y^2+z^2)^{\frac{3}{2}}}\right]\left[(\dot
x )^2+(\dot y )^2+(\dot z )^2\right]\right\}.\eqf From it the
geodesic equations of motions are \eqia \ddot
t\left[1+\frac{kz}{(x^2+y^2+z^2)^{\frac{3}{2}}}\right]&=&
\frac{3kz\dot t(x\dot x+y\dot y+z\dot z)}
{(x^2+y^2+z^2)^{\frac{5}{2}}}-\frac{k\dot z\dot t}{(x^2+y^2+z^2)^{\frac{3}{2}}},\\
\ddot
x\left[1-\frac{kz}{(x^2+y^2+z^2)^{\frac{3}{2}}}\right]&=&\frac{k\dot
x\dot z }{(x^2+y^2+z^2)^{\frac{3}{2}}}-\frac{3k\dot x z(x\dot
x+y\dot y+z\dot z )}{(x^2+y^2+z^2)^{\frac{5}{2}}}+\nonumber\\
&+&\frac{3kzxc^2(\dot
t)^2}{2(x^2+y^2+z^2)^{\frac{5}{2}}}+\frac{3kzx[(\dot x)^2+(\dot y
)^2+(\dot z)^2]}{2(x^2+y^2+z^2)^{\frac{5}{2}}},\lb{bestia1}\\
\ddot
y\left[1-\frac{kz}{(x^2+y^2+z^2)^{\frac{3}{2}}}\right]&=&\frac{k\dot
y\dot z }{(x^2+y^2+z^2)^{\frac{3}{2}}}-\frac{3k\dot y z(x\dot
x+y\dot y+z\dot z )}{(x^2+y^2+z^2)^{\frac{5}{2}}}+\nonumber\\
&+&\frac{3kzyc^2(\dot
t)^2}{2(x^2+y^2+z^2)^{\frac{5}{2}}}+\frac{3kzy[(\dot x)^2+(\dot y
)^2+(\dot z)^2]}{2(x^2+y^2+z^2)^{\frac{5}{2}}},\lb{bestia2}\\
\ddot z\left[1-\frac{kz}{(x^2+y^2+z^2)^{\frac{3}{2}}}\right]&=&
\frac{k(\dot z)^2}{(x^2+y^2+z^2)^{\frac{3}{2}}}-\frac{3kz\dot
z(x\dot x+y\dot y+z\dot z)
}{(x^2+y^2+z^2)^{\frac{5}{2}}}+\frac{3kz^2 c^2(\dot
t)^2}{2(x^2+y^2+z^2)^{\frac{5}{2}}}-\nonumber\\
&-&\frac{kc^2(\dot
t)^2}{2(x^2+y^2+z^2)^{\frac{3}{2}}}+\frac{3kz^2[(\dot x )^2+(\dot
y )^2+(\dot
z)^2]}{2(x^2+y^2+z^2)^{\frac{5}{2}}}-\nonumber\\
&-&\frac{k[(\dot x )^2+(\dot y )^2+(\dot
z)^2]}{2(x^2+y^2+z^2)^{\frac{3}{2}}}.\lb{bestia3}\eqfa Note  that
\rfrs{bestia1}{bestia3} for $k=0$ reduce to the ordinary $\ddot
x=\ddot y=\ddot z=0$ of the flat spacetime case.

It is interesting to note that the relativistic acceleration
induced by the rotating sphere is non central, has some terms
which are quadratic in the velocity $\bm v$ of the test particle
and depends on the inverse of the third power of the distance.
Such features could be important in view of a possible
experimental investigation.
Notice that the classical Newtonian acceleration of the sphere
turns out to be far smaller than the relativistic acceleration.
For a sphere with $\omega=4.33\times 10^4$ rad s$^{-1}$,
$R=2.5\times 10^{-2}$ m, $M=1.11$ kg and particles with
$v=3.9\times 10^3$ m s$^{-1}$, as for, say, thermal neutrons, we
have that the ratio of the relativistic acceleration
$\sim\frac{kv^2}{r^3}$ to the Newtonian one $\frac{GM}{r^2}$ is
\eqi\frac{kv^2}{GM r}=\frac{1.4\times 10^5\ {\rm m} }{r}.\eqf
Moreover, while\footnote{Notice that the radius of curvature of
the particle trajectory induced by the Newtonian acceleration of
gravity of the spinning sphere would be 4.86$\times 10^{19}$ m,
i.e. the particles would fly along straight lines.}
$\frac{GM}{r^2}\sim 10^{-9}$ m s$^{-2}$ for $r=5\times 10^{-2}$ m,
$\frac{kv^2}{r^3}\sim 8\times 10^{-4}$ m s$^{-2}$ for $r=5\times
10^{-2}$ m. Of course, if we would think about some experiments on
the Earth's surface, its acceleration of gravity $g=9.86$ m
s$^{-2}$ should be accounted for.
\subsection{An Earth laboratory free fall scenario}
\begin{figure}[ht!]
\begin{center}
\includegraphics*[width=10cm,height=8cm]{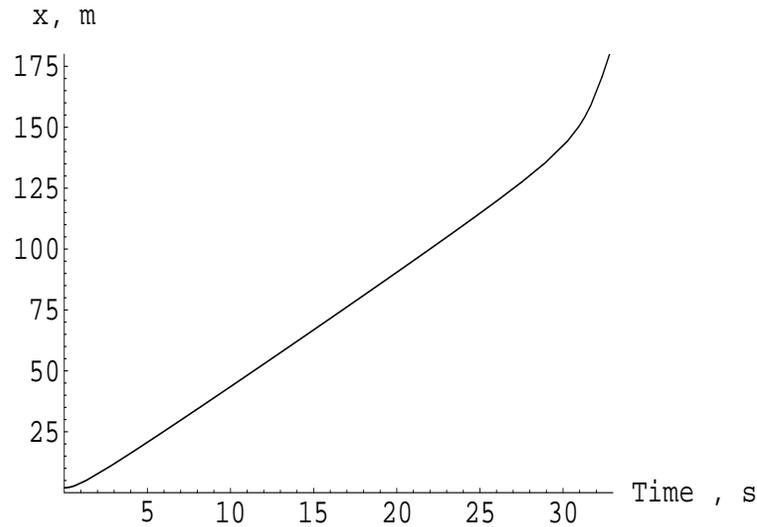}
\end{center}
\caption{\footnotesize Time evolution of the $x$ coordinate of a
point mass freely falling in the Earth's gravitational field and
in the local gravitational field of a spinning sphere with
$\omega=6.28\times 10^3$ rad s$^{-1}$, $R=1$ m. The initial
conditions are $x(0)=y(0)=2$ m, $z(0)=120$ m, $\dot x(0)=\dot
y(0)=\dot z(0)=0$. The origin of the coordinates is the center of
the sphere and the $z=0$ plane is its equatorial plane.}
\label{figura2}
\end{figure}

\begin{figure}[ht!]
\begin{center}
\includegraphics*[width=10cm,height=8cm]{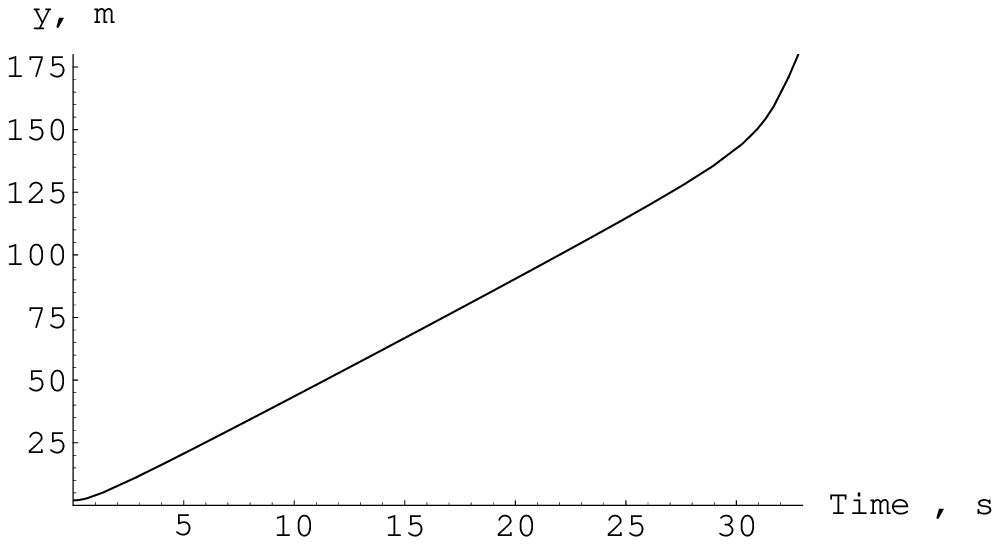}
\end{center}
\caption{\footnotesize Time evolution of the $y$ coordinate of a
point mass freely falling in the Earth's gravitational field and
in the local gravitational field of a spinning sphere with
$\omega=6.28\times 10^3$ rad s$^{-1}$, $R=1$ m. The initial
conditions are $x(0)=y(0)=2$ m, $z(0)=120$ m, $\dot x(0)=\dot
y(0)=\dot z(0)=0$. The origin of the coordinates is the center of
the sphere and the $z=0$ plane is its equatorial plane.}
\label{figura3}
\end{figure}

\begin{figure}[ht!]
\begin{center}
\includegraphics*[width=10cm,height=8cm]{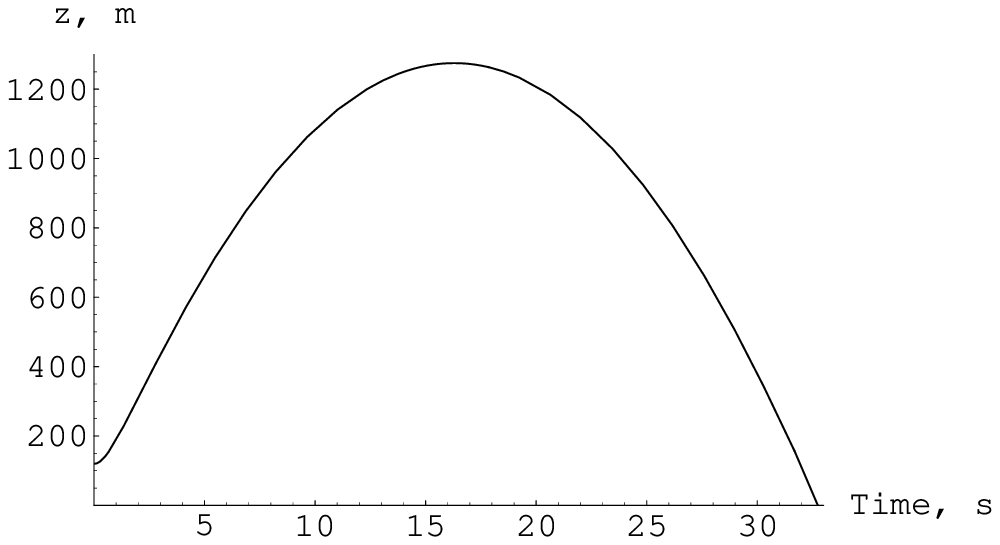}
\end{center}
\caption{\footnotesize Time evolution of the $z$ coordinate of a
point mass freely falling in the Earth's gravitational field and
in the local gravitational field of a spinning sphere with
$\omega=6.28\times 10^3$ rad s$^{-1}$, $R=1$ m. The initial
conditions are $x(0)=y(0)=2$ m, $z(0)=120$ m, $\dot x(0)=\dot
y(0)=\dot z(0)=0$. The origin of the coordinates is the center of
the sphere and the $z=0$ plane is its equatorial plane. Note that,
according to Newtonian gravity, $t(z=0)=4.9$ s. } \label{figura4}
\end{figure}

In order to examine a concrete scenario at Earth laboratory scale,
we have numerically solved\footnote{In doing so we have posed
$\left(1\pm\frac{kz}{r^3}\right)=1$ and $\dot t=1$. We have also
checked that, as it could be expected, the results of the
integration do not change if the full expressions of the geodesic
equations are retained. } \rfrs{bestia1}{bestia3} with the
software $MATHEMATICA$ for two different scenarios. In the first
one we have assumed $k=3.67\times 10^{-9}$ m$^2$, which
corresponds to a spinning sphere with angular velocity
$\omega=6.28\times 10^3$ rad s$^{-1}$ and radius $R=1$ m, with the
initial conditions $x(0)=y(0)=2$ m, $z(0)=120$ m, $\dot x(0)=\dot
y(0)=\dot z(0)=0$. In order to account for the Earth's gravity
acceleration $g$ we have added $-g$ to the right-hand side of
\rfr{bestia3} The interesting results are plotted in Figure
\ref{figura2}-Figure \ref{figura4}. It can be noted that, while in
the classical case of free fall $x$ and $y$ remain constant and
$z$ vanishes after\footnote{It is exactly the result it can be
obtained by putting $k=0$ in solving the equations.} $t=4.94$ s,
in this case the falling body starts going upwards acquiring a
displacement from the vertical of a large amount, after almost 15
s it inverts its motion downwards and finally it reaches the
reference plane $z=0$ in 33 s. It seems a very strange behavior.

\begin{figure}[ht!]
\begin{center}
\includegraphics*[width=10cm,height=8cm]{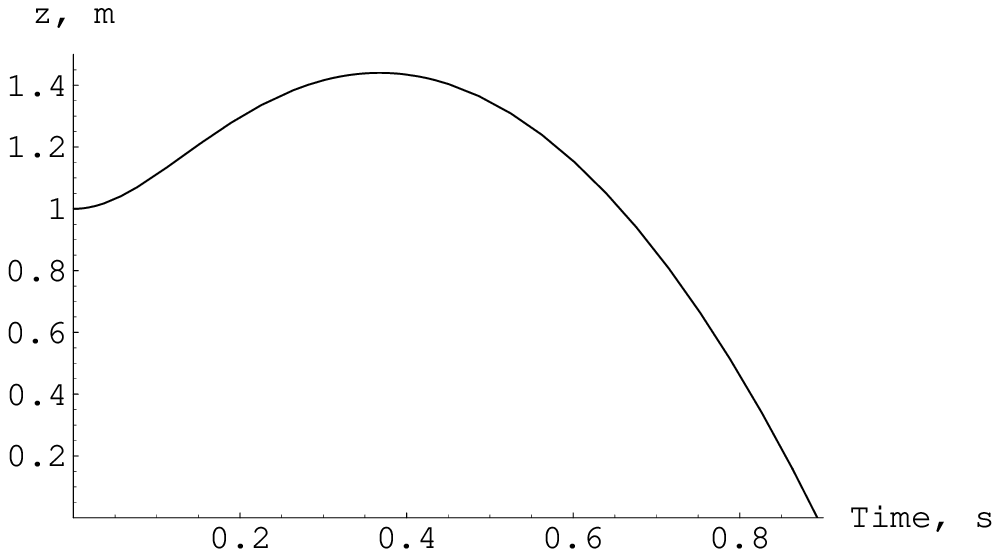}
\end{center}
\caption{\footnotesize Time evolution of the $z$ coordinate of a
point mass freely falling in the Earth's gravitational field and
in the local gravitational field of a spinning sphere with
$\omega=10^2$ rad s$^{-1}$, $R=2\times 10^{-1}$ m. The initial
conditions are $x(0)=y(0)=5\times 10^{-1}$ m, $z(0)=1$ m, $\dot
x(0)=\dot y(0)=\dot z(0)=0$. The origin of the coordinates is the
center of the sphere and the $z=0$ plane is its equatorial plane.
Note that, according to Newtonian gravity, $t(z=0)=0.4$ s. }
\label{figura6}
\end{figure}

\begin{figure}[ht!]
\begin{center}
\includegraphics*[width=10cm,height=10cm]{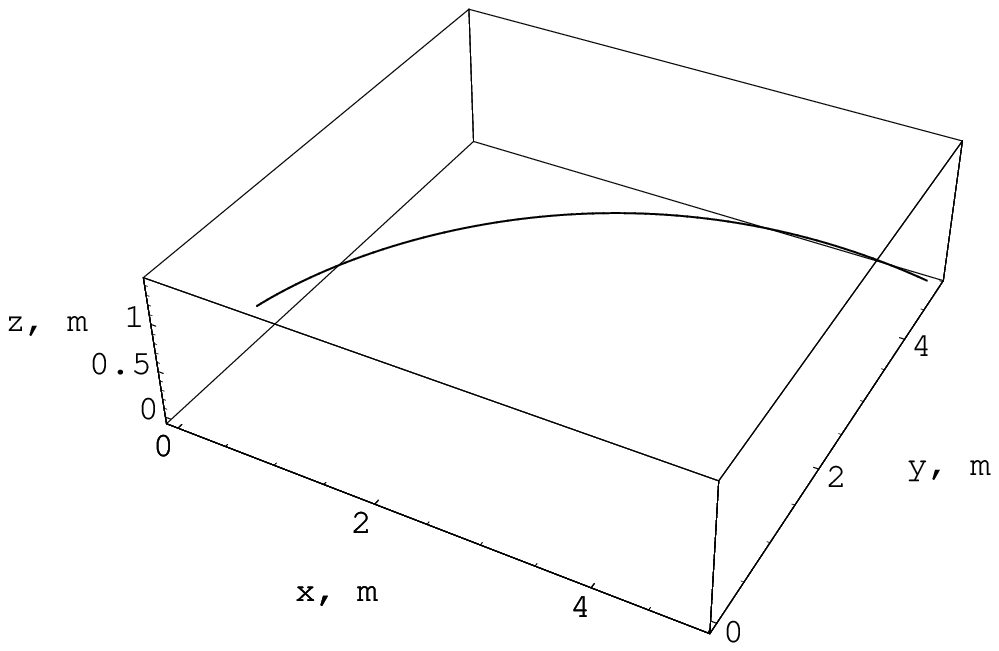}
\end{center}
\caption{\footnotesize Spatial trajectory followed by a point mass
freely falling in the Earth's gravitational field and in the local
gravitational field of a spinning sphere with $\omega= 10^2$ rad
s$^{-1}$, $R=2\times 10^{-1}$ m. The initial conditions are
$x(0)=y(0)=5\times 10^{-1}$ m, $z(0)=1$ m, $\dot x(0)=\dot
y(0)=\dot z(0)=0$. The origin of the coordinates is the center of
the sphere and the $z=0$ plane is its equatorial plane.}
\label{figura7}
\end{figure}

In the second scenario, for a simple sphere with just
$\omega=10^2$ rad s$^{-1}$ and $R=2\times 10^{-1}$ m,
corresponding to $k=1.5\times 10^{-15}$ m$^2$, and
$x(0)=y(0)=5\times 10^{-1}$ m, $z(0)=1$ m and $\dot x(0)=\dot
y(0)=\dot z(0)=0$ it would be possible to observe\footnote{It can
be shown that in both the cases considered here the speeds reached
by the freely falling bodies are nonrelativistic.} a notable
deviation with respect to the vertical and an increment of the
time required to reach the $z=0$ plane, as shown in Figure
\ref{figura6} and Figure \ref{figura7}. It should be noted that
the mechanical parameters of the spinning sphere of this scenario
are quite common and easy to obtain, so that it should be possible
to observe the exotic effects described here in many ordinary
situations.
\subsection{A gravitational shielding effect?}
Let us consider now the case of a body suspended upon the rotating
source at $x=y=0,$ $z=h$. It can be shown that the weight of the
suspended body is reduced by the local gravitational field of the
spinning disk. Indeed, from \rfrs{bestia1}{bestia3} it can be
noted that there are no in-plane, tangential components of the
relativistic acceleration while the Earth's gravity acceleration
along the vertical is reduced according to \eqi\frac{d^2 z}{d
t^2}=-g+\frac{kc^2}{h^3}.\lb{schermo1}\eqf For a generic position
of the sample there is also a tangential part of the acceleration:
indeed we have \eqia \frac{d^2 x}{d t^2}&=& \frac{3kc^2x_0
z_0}{2(x_0^2+y_0^2+z_0^2)^{\frac{5}{2}}},\\
\frac{d^2 y}{d t^2}&=& \frac{3kc^2y_0
z_0}{2(x_0^2+y_0^2+z_0^2)^{\frac{5}{2}}},\\
\frac{d^2 z}{d
t^2}&=&-g+\frac{kc^2}{2}\left[\frac{3z_0^2}{(x_0^2+y_0^2+z_0^2)^{\frac{5}{2}}}-
\frac{1}{(x_0^2+y_0^2+z_0^2)^{\frac{3}{2}}}\right].\lb{schermo3}\eqfa
For a spinning sphere with $\omega=10^2$ rad s$^{-1}$ and
$R=2\times 10^{-1}$ m, i.e. $k=1.5\times 10^{-15}$ m$^2$ and a
body suspended over the disk with $x=y=0$ and $z=h=3$ m, the
weight-reducing contribution would be 4.96 m s$^{-2}$, which is a
very large figure. Note also that the relativistic additional
forces of \rfrs{schermo1}{schermo3} are independent of the speed
of light $c$.

Let us consider now the case of a disk with inner radius $R_1$,
outer radius $R_2$, thickness $l$ and uniform density $\rho$
placed in rotation around an axis orthogonal to its plane assumed
as $z$ axis. In this case \eqi a =\frac{\left(R_2^4
-R_1^4\right)\omega}{2c\left(R_2^2-R_1^2\right)}.\eqf For
$\omega=5\times 10^3$ rpm=$5.23\times 10^2$ rad s$^{-1}$,
$R_1=4\times 10^{-2}$ m, $R_2=1.375\times 10^{-1}$ m we have
$k=C_0 a^2=C_0\times 3.2\times 10^{-16}$ m$^2$. For a body
suspended over the disk with $x=y=0$ and $z=h=3$ m the correction
to the Earth's gravity acceleration amounts to $C_0\times 1.06$ m
s$^{-2}$, i.e. $C_0\times$10$\%$ of $g$.

It is interesting a comparison with the famous and controversial
antigravity experiment by E.E. Podkletnov  \cite{pod97}. In it it
was reported that a high-temperature $YBa_2 Cu_3O_{7-x}$ bulk
ceramic superconductor with composite structure has revealed weak
shielding properties against gravitational force while in a
levitating state at temperatures below 70 $K$. A toroidal disk
with an outer diameter of $2.75\times 10^{-1}$ m and a thickness
of $1\times 10^{-2}$ m was prepared using conventional ceramic
technology in combination with melt-texture growth. Two solenoids
were placed around the disk in order to initiate the current
inside it and to rotate the disk around its central axis. Material
bodies placed over the rotating disk initially demonstrated a
weight loss of 0.3-0.5$\%$. Moreover, the air over the cryostat in
which the apparatus was enclosed began to rise slowly toward the
ceiling. Particles of dust and smoke in the air made the effect
clearly visible. Interestingly, the boundaries of the flow could
be seen clearly and corresponded exactly to the shape of the
toroid. When the angular velocity of the disk was slowly reduced
from 5,000 rpm to 3,500 rpm by changing the current in the
solenoids, the shielding effect became considerably higher and
reached 1.9-2.1$\%$ at maximum. Moreover, the effective weight
loss turned out to be independent of the height of the suspended
bodies over the disk. Finally, the shielding effect was present
even in absence of rotation ranging from 0.05$\%$ to 0.07$\%$.

It is evident that the phenomenology described in \cite{pod97}
cannot be accounted for by the general relativistic phenomena
considered here.

It could be interesting to mention that a sort of genuine
antigravitational effect does exist in General Relativity. It is
related to the behavior of a test particle which moves along the
rotational axis of a naked singularity in the Kerr metric: it is
the so called rotational paradox \cite{cohfdf84, fdf75, fdf93}.
\section{Conclusions}
In this paper we have derived some features of the motion of test
particles in a spacetime metric represented by an approximated
solution of the Einstein equations for a weakly gravitating
spinning object in which the mass effects are negligible with
respect to the rotational effects. These conditions are well
satisfied at laboratory scale. Since it is doubtful that such
metric could really describe the gravitational field of a material
body, such calculations could also be viewed as an attempt to shed
light on its validity by looking for possible strange or non
existent observable consequences.

Then, we have derived the geodesic equations of motion for a
massive test particle. Some rather puzzling features have been
found in the allowed geodesic motions. For example, neither
circular motions in the equatorial plane of the source nor
spherical motions would be permitted. We have numerically solved
the geodesic equations for a pair of particular choices of initial
conditions representing, in classical Newtonian mechanics, a
vertical free-fall motion in a possible laboratory scenario. For a
realistic choice of the mechanical parameters of the sphere the
investigated effects would be quite measurable. We have assumed
$C_0=\frac{50}{3}\pi$. We have found that the rotation of the
central source would induce a sort of antigravity effect with an
increment of the time required to reach the ground and also a
deviation of the trajectory from the local vertical. We have also
noted that the weight of a massive body turns out to be reduced by
the acceleration considered here. The crucial point is that for a
realistic and rather common choice of the mechanical parameters of
the central rotating sphere all such effects seem to be very large
in magnitude: nothing similar to them has ever been observed.

As a conclusion, the obtained results in this paper might be
considered as a further insight against any real physical
significance of the approximated metric considered here.
\section*{Appendix: The second order powers of the rotation in the PPN equations of motion}
The equations of motion for a particle orbiting a finite sized
spherical extended body with constant density in the PPN formalism
can be found in \cite{nor94}. From them the contribution of the
square of the angular velocity of the central body to the
particle's acceleration can be extracted \cite{pet97}: it consists
of two radial terms and a third term directed along the spin axis
of the rotating mass. For a spherical rotating body with the $z$
axis directed along its spin axis General Relativity yields\eqia
\ddot x &=& \frac{G}{c^2}M
R^4\omega^2\left[\frac{6xz^2}{7(x^2+y^2+z^2)^{\frac{7}{2}}}-\frac{6x}{35(x^2+y^2+z^2)^{\frac{5}{2}}}-
\frac{3x}{5R^2(x^2+y^2+z^2)^{\frac{3}{2}}}\right],\\\lb{pet1}
\ddot y &=& \frac{G}{c^2}M
R^4\omega^2\left[\frac{6yz^2}{7(x^2+y^2+z^2)^{\frac{7}{2}}}-\frac{6y}{35(x^2+y^2+z^2)^{\frac{5}{2}}}-
\frac{3y}{5R^2(x^2+y^2+z^2)^{\frac{3}{2}}}\right],\\
\ddot z &=& \frac{G}{c^2}M
R^4\omega^2\left[\frac{6z^3}{7(x^2+y^2+z^2)^{\frac{7}{2}}}-\frac{18z}{35(x^2+y^2+z^2)^{\frac{5}{2}}}-
\frac{3z}{5R^2(x^2+y^2+z^2)^{\frac{3}{2}}}\right].\lb{pet3}\eqfa
The second order powers of the rotation have here been interpreted
as arising from the rotational energy of the central body. These
accelerations come from the potentials evaluated for the $g_{00}$
portion of the metric and as such originate from the curvature of
the spacetime. They are, therefore, related to the energy and do
not arise from a second order power of the angular momentum of the
body.

Note that in \rfrs{pet1}{pet3} the factor
$\frac{G}{c^2}=7.42\times 10^{-28}$ m kg$^{-1}$ is present, so
that their effects at laboratory scale are completely negligible.
\section*{Acknowledgements}
I am grateful to L. Guerriero for his support while at Bari.
Special thanks to B. Mashhoon for useful and important discussions
and clarifications and to F. de Felice for the explanations of the
rotational paradox.

\end{document}